\documentclass[%
 reprint,
superscriptaddress,
showpacs,preprintnumbers,
 amssymb,
 aps,
]{revtex4-1}

\usepackage{graphicx} 
\graphicspath{{Figures/}}

\usepackage{dcolumn}

\usepackage{bm}        
\usepackage{amssymb}   
\usepackage{mathtools}  
\usepackage{mathrsfs}
\usepackage{multirow}
\usepackage{booktabs}
\usepackage{subcaption}
\usepackage{ragged2e}
\DeclareCaptionJustification{justified}{\justifying}
\usepackage{soul}
\usepackage[inline]{enumitem}
\usepackage[svgnames, table]{xcolor}
\usepackage[colorlinks=true, citecolor=Grey,
            linkcolor=DarkOrchid, urlcolor=ForestGreen]{hyperref}

\usepackage{cleveref}

\def\be{\begin{equation}}
\def\ee{\end{equation}}
\def\ba{\begin{eqnarray}}
\def\ea{\end{eqnarray}}

\begin{document}

\newcommand{\R}[1]{\textcolor{Red}{#1}} 
\newcommand{\G}[1]{\textcolor{ForestGreen}{#1}} 
\newcommand{\B}[1]{\textcolor{MediumOrchid}{#1}} 
\newcommand{\K}[1]{\textcolor{YellowOrange}{#1}} 
\newcommand{\C}[1]{\textcolor{DodgerBlue}{#1}} 


\title{A phase-sensitive optomechanical amplifier for quantum noise reduction in laser interferometers}


\author{Yuntao Bai}
\affiliation{
  Walter Burke Institute for Theoretical Physics,
  California Institute of Technology
}
\author{Gautam Venugopalan}
\email[E-mail me at: ]{gautam@caltech.edu}
\affiliation{%
 LIGO Laboratory, California Institute of Technology
}
\author{Kevin Kuns}
\affiliation{%
 LIGO Laboratory, Massachussetts Institute of Technology
}
\affiliation{%
 LIGO Laboratory, California Institute of Technology
}

\author{Christopher Wipf}
\affiliation{%
 LIGO Laboratory, California Institute of Technology
}
\author{Aaron Markowitz}
\affiliation{%
  LIGO Laboratory,
  California Institute of Technology
}
\author{Andrew R. Wade}
\affiliation{%
  LIGO Laboratory,
  California Institute of Technology
}
\author{Yanbei Chen}
\affiliation{
  Walter Burke Institute for Theoretical Physics,
  California Institute of Technology
}
\author{Rana X Adhikari}
\affiliation{%
  LIGO Laboratory,
  California Institute of Technology
}





\date{\today}

\begin{abstract}
The sensitivity of future gravitational wave interferometers is expected to be limited throughout the detection band by quantum vacuum fluctuations, which can be reduced by applying quantum optics techniques such as squeezed vacuum injection. However, decoherence caused by optical losses in the readout chain will severely limit the effectiveness of such schemes. It was proposed that effect of losses in the final stage of detection can be mitigated by a phase-sensitive amplifier placed in between the output port of the interferometer and the photodetector. In this paper, we propose to implement such amplification using an optomechanical device, study some of its practical limitations, and discuss its applicability to next-generation gravitational-wave detectors.

\end{abstract}


\maketitle



\begin{figure}
  \includegraphics[width=\columnwidth]{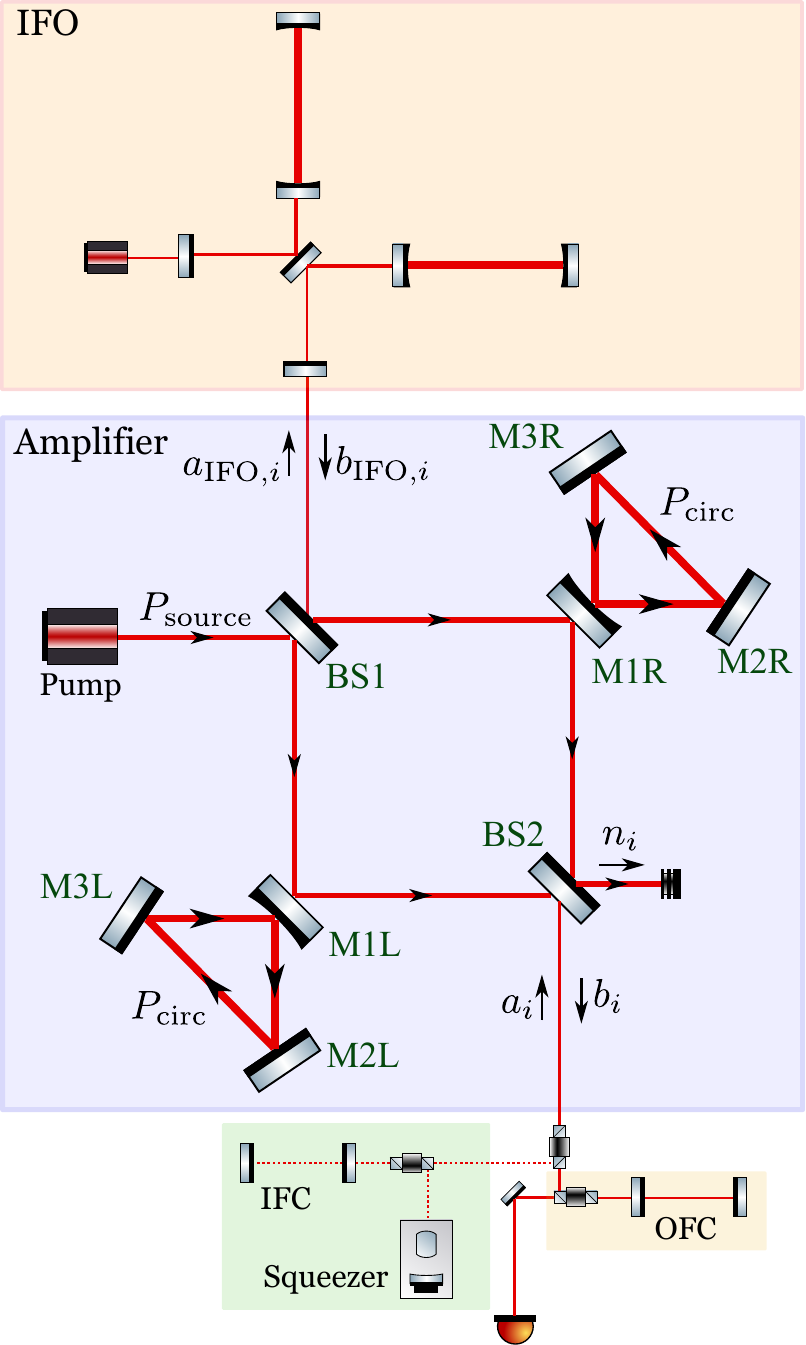}
\caption{Optical layout for the Mach-Zehnder amplifier (shaded in light purple) installed between the anti-symmetric port of the IFO (shaded in light red) and the readout chain, with arrows indicating the direction of the amplifier pump. \textit{The various subsystems are not drawn to scale}. IFC: input filter cavity; OFC: output filter cavity; BS1, BS2: 50/50 beam splitter; M1R, M1L: highly reflecting mirror; M2R, M2L, M3R, M3L: perfectly reflecting mirror}
  \label{fig:MZ}
\end{figure}

\section{Introduction}
\label{sec:intro}

The first direct detection of gravitational wave (GW) signals from outer space was achieved by the LIGO collaboration in 2015~\cite{GW150914}, marking the beginning of gravitational wave astronomy. Since then, multiple other detections have been announced~\cite{abbott2016gw151226,scientific2017gw170104,abbott2017gw170814,abbott2017gw170817,abbott2017gw170608,ligo2018gwtc}. These signals were produced by some of the most violent events in the universe---the coalescence of compact objects such as black holes and neutron stars. They provide decisive tests of general relativity in the strong-gravity regime, may give clues about the rich nuclear physics within the ultra-dense cores of neutron stars, and most importantly have the potential to teach us many unexpected lessons about the universe. Some of the most interesting physics of these phenomena appear during and after these merger events, but the details of the signal wave forms detected so far have been masked by the photon shot noise in the interferometers' readout. 

Despite the limitations imposed by the quantum nature of light, shot noise is not a fundamental limit of nature, and therefore can be mitigated by clever optical techniques. In the 1980's, Carlton Caves showed that the sensitivity of shot noise-limited interferometers can be improved by injecting `squeezed vacuum'~\cite{Caves:1981}. This paved the way for `quantum non-demolition' technologies for GW detection, such as those explored in~\cite{KLMTV:2001}. Since then, squeezed vacuum injection has been implemented successfully both at GEO600~\cite{vahlbruch2010geo, sqzGEO:2011} and at LIGO~\cite{H1:SQZ}. However, the extent to which sensitivity can be improved by this technique is limited by losses incurred within the interferometer optics, leading to decoherence of the squeezed field.

One way of overcoming losses in the final stage of photodetection, as proposed by Caves ~\cite{Caves:1981}, is to pre-amplify the quadrature field containing the GW signal, using a quantum-limited phase-sensitive amplifier, before reading it out with a photodetector.  This idea was more recently discussed by Knyazev et al.~\cite{knyazev2018quantum,knyazev2019overcoming}.
In this work, we propose an optical layout that serves as an ultra-low noise, high gain, and phase-sensitive optomechanical amplifier for the GW signal, as shown in~\Cref{fig:MZ}.  We describe how this amplifier can mitigate the effect of optical losses in the readout chain and therefore let us listen to the universe in higher fidelity. Significant progress has been made toward demonstrating optomechanical parametric amplification in the gravitational-wave band, in order to generate squeezed vacuum~\cite{Corbitt2005,cripe2019measurement}.  The amplifier we propose and analyze here will be another application of such devices.

This paper is organized as follows:
we begin in~\Cref{sec:voyager} with a qualitative discussion of the role of quantum noise mitigation and the need for an amplifier in the context of LIGO Voyager~\cite{Voyager:Inst,Voyager:Science}, a next generation upgrade to LIGO.
Then, in~\Cref{sec:opto}, we explain the physics of optomechanical amplification, and provide some simple formulae whose detailed derivations are postponed to~\Cref{app:ring}. We then propose, in the same section, an optical layout for the amplifier. In~\Cref{sec:noise}, we discuss noise sources within the amplifier, which necessarily limit the amplifier's performance. The effectiveness of the amplifier is further limited by noise sources within the main interferometer (IFO), the most serious of which are discussed in~\Cref{sec:IFO_noise}. Finally, in~\Cref{app:mirror_mass} we discuss some parameter choices for the proposed design, and in~\Cref{app:20dB} we discuss prospects for a more ambitious quantum amplifier.

\section{Squeezed Vacuum Injection in LIGO Voyager}
\label{sec:voyager}
Voyager is a planned cryogenic upgrade to the terrestrial gravitational wave detectors~\cite{Voyager:Inst}.
The primary sensitivity improvement is realized by operating with silicon test masses and amorphous silicon optical coatings at 123\,K. At this temperature, the coefficient of linear thermal expansion of silicon, $\alpha$, vanishes, which drives down phase noise imparted by thermally-driven fluctuations in the interferometer.
Design studies and early R\&D suggest that the sensitivity of Voyager will be limited throughout the detection band of 10\,--\,4000\,Hz by quantum noise, provided that the Brownian noise of the optical coating is sufficiently mitigated (see~\Cref{sec:IFO_noise} for more details).
To reduce the quantum noise, squeezed vacuum is injected via the anti-symmetric/dark port (see \Cref{fig:MZ}) ~\cite{Caves:1981}.
The rotation angle of the noise ellipse of the squeezed vacuum required for broadband sensitivity improvement is frequency-dependent. This frequency-dependence is realized by installing a filter cavity along the injection path~\cite{KLMTV:2001}.
We refer to this as the `input filter cavity' (IFC), which is illustrated in~\Cref{fig:MZ}.

In principle, injecting more strongly squeezed vacuum improves the sensitivity monotonically.
In practice, however, the improvement is limited by optical losses in various parts of the interferometer.
At each point where loss occurs, the coherence of the squeezed vacuum is degraded. The relative importance of loss at various ports in the core interferometer is shown in \Cref{fig:ifoNoises}.
It is anticipated that optical losses in the Voyager arm cavities will be reduced to the tens of ppm level per round-trip, due to ultra high quality optical surfaces and coatings.
However, the readout chain, consisting of an Output Faraday Isolator (OFI), Output Mode Cleaner (OMC), and balanced homodyne detection with photodetectors that have imperfect quantum efficiency, is expected to introduce significant loss, possibly at the level of $10\%$. Mitigating the readout loss directly would require substantial effort to improve multiple pieces of technology. However, requirements on these losses can be relaxed by pre-amplifying the GW signal, provided that the amplifier
(a) has ultra-low noise,
(b) has high gain in the frequency band of the GW signal, and
(c) is phase-sensitive and therefore immune to the quantum mechanical limits of phase-insensitive amplifiers described in~\cite{Caves:1982}.
In the remainder of this paper, we show how an optomechanical amplifier installed between the GW interferometer and the readout chain achieves all three properties, and discuss the resulting impact on detector sensitivity.

\section{Optomechanics for phase-sensitive amplification}
\label{sec:opto}

In this section, we begin by reviewing the physics of optomechanical amplification in \Cref{sec:amp}. Then we propose an optical layout and design parameters for the amplifier in~\Cref{sec:MZ}, and explain how it should be incorporated in the main interferometer.

We analyze optomechanical interactions using the two-photon formalism developed by Caves and Schumaker~\cite{Caves:1985_1, Caves:1985_2}, and reviewed in Sec. II. A. of~\cite{KLMTV:2001}. In particular, we use the notation and Fourier transform convention followed in the latter. We let $\omega_0/2\pi$ denote the carrier frequency of the laser beam in the main interferometer, and $\Omega/2\pi\lesssim 4\,{\rm kHz}$ denote the signal sideband frequency. 

\subsection{Optomechanical amplification}
\label{sec:amp}

Optomechanical amplification~\cite{PhysRevA.85.013812} is a process by which a signal beam is amplified (i.e. anti-squeezed) via the radiation pressure coupling between the optical field and the mechanical modes of a suspended mirror. To enhance the coupling, the signal beam can be applied to the mirror together with a co-propagating pump beam. We decompose the signal into two quadratures in the usual way, referring to the quadrature in phase with the IFO pump amplitude as the `amplitude quadrature', and the orthogonal quadrature as the `phase quadrature'. Signal and pump fields interfere to produce amplitude and phase fluctuations in the light incident on the mirror. Incident amplitude fluctuations exert radiation pressure, which displaces the mirror, thus modulating the phase of the reflected light  (eq.~\ref{eq:forward_approx}, \ref{eq:ring_K}). In the limit of a strong pump and low mirror mass, the induced phase fluctuations on reflection may be much larger than the amplitude fluctuations of the incident signal. Since this process only amplifies one quadrature, it is phase-sensitive and we reiterate that it is not subject to the quantum mechanical noise limits of phase-insensitive amplifiers~\cite{Caves:1982}.

Before discussing Voyager, let us provide a simple explanation why phase sensitive amplification, proposed by Caves, can improve robustness against losses in the detection chain.  Suppose the $b_1$ quadrature carries signal $h$ and squeezed noise, $e^{-r} a_1$,
\begin{equation}
    b_1 = h + e^{-r} a_1.
\end{equation}

This corresponds to a signal-referred noise spectrum of $S_h =e^{-2r}$ (see \cite{KLMTV:2001} for details of the two-photon formalism).   If we were to detect this quadrature with losses, $\epsilon$, in the readout chain, we will be detecting

\begin{equation}
    b_1^D = \sqrt{1-\epsilon} \left[h + e^{-r} a_1\right] +\sqrt{\epsilon}n_1\,,
    \label{eq:lossyQuad}
\end{equation}

which leads to

\begin{equation}
    S_h^{\rm loss} = e^{-2r}+\epsilon\,.
    \label{eq:lossySNR}
\end{equation}

In \cref{eq:lossyQuad}, $n_1$ and $a_1$ denote unsqueezed vacuum. This simplified scenario highlights how losses in the detection chain, $\epsilon$, limit the sensitivity enhancement we are able to achieve by injecting squeezed vacuum.  Now, suppose we feed $b_1$ into a phase-sensitive amplifier, which allows noise-free linear amplification by $G$, before feeding the amplified quadrature field into the \emph{same} detection chain. Then, we will have
\begin{equation}
    b_1^{D,\,{\rm amp}} = \sqrt{1-\epsilon} \,G\left[h + e^{-r} a_1\right] +\sqrt{\epsilon}n_1\,,
\end{equation}

with a signal-referred noise spectrum of

\begin{equation}
    S_h^{\rm loss,\,amp}=e^{-2r}+\frac{\epsilon}{G^2}.
\end{equation}

In this way, the noise power due to losses in the detection process is suppressed by $G^2$. Note that in this process, the orthogonal quadrature $b_2$, which carries no signal, is suppressed by $G$.

\subsection{The Mach-Zehnder Amplifier}
\label{sec:MZ}

For the optical layout of the amplifier, we propose a Mach-Zehnder (MZ) configuration to be installed between the anti-symmetric port of the main LIGO interferometer and the readout chain, as shown in~\Cref{fig:MZ}. The topology consists of two input ports, with the output signal from the LIGO interferometer (labeled as $b_{{\rm IFO,}\, i}$) injected into one, and a pump (labeled as $P_{{\rm source}}$) with the same carrier frequency injected into the other. The pump and signal are combined at a 50/50 beamsplitter (BS1) and split into two beams, with each beam directed to a separate triangular ring cavity. In each ring, the beating between the pump and signal produces the optomechanical amplification discussed in~\Cref{sec:amp}. To enhance the gain, the mirrors of the ring are designed to weigh as little as possible, and the cavity length is locked to have the pump field be resonant, in order to achieve high circulating power. Finally, the output beams of the ring cavities are recombined at a second 50/50 beamsplitter (BS2), with the amplified signal (labeled as $b_i$) measured at one port and the strong pump field dumped at the other.


We now present the input-output relations for the MZ amplifier, where $b_{{\rm IFO,}\, i}$ denote the field amplitudes for the input quadratures to the amplifier, and $b_i$ the field amplitudes for the output quadratures, as shown in~\Cref{fig:MZ}. The normalization of the field amplitudes is defined in (6) of~\cite{KLMTV:2001}.

We let $R_{\rm A},T_{\rm A}$ denote the power reflectivity and transmissivity respectively of the M1L \& M1R mirrors, and $L_A$ denote the round-trip length of each ring. A more exact calculation in the lossless limit is given in~\Cref{app:ring}, where additional assumptions are explained. Here we simply present the result in the limit $\Omega L_A/c\ll1$ and $T_{\rm A}\ll 1$, giving
\begin{align}
\label{eq:forward_approx}
\!\begin{multlined}[t]
\begin{pmatrix}
b_1\\
b_2
\end{pmatrix}
=
e^{2i\eta}\begin{pmatrix}
1 & 0 \\
-\mathcal{K}_{\rm A} & 1
\end{pmatrix}
\begin{pmatrix}
b_{\rm IFO,\, 1}\\
b_{\rm IFO,\, 2}
\end{pmatrix} \\ +
\sqrt{\frac{32 \omega_0 P_{\rm circ}}{\hbar c^2}} \frac{1}{T_{\rm A}}
\begin{pmatrix}
0 \\
1
\end{pmatrix} \xi,
\end{multlined}
\end{align}
where
\ba\label{eq:ring_K}
\mathcal{K}_{\rm A}&=&\frac{4}{T_{\rm A}\left[
1+\left(
\Omega/\gamma_{\rm A}\right)^2
\right]}
\kappa_{\rm A}
,
\nonumber
\\
\kappa_{\rm A}&=&-\frac{18\omega_0P_{\rm circ}}{c^2}\chi_{\rm A},
\nonumber
\\
\eta&=&\arctan\left(\Omega/\gamma_{\rm A}\right),\qquad
\gamma_{\rm A}=\frac{cT_{\rm A}}{2L_{\rm A}},
\ea
with $\gamma_{\rm A}$ as the cavity pole frequency, $P_{\rm circ}$ as the power circulating in each ring, $\chi_{\rm A}$ as the mechanical susceptibility of the movable mirrors which for a mirror of mass $m_{\rm A}$ suspended as a lossless pendulum of natural frequency $\Omega_0/2\pi$ is given by
\ba
\chi_{\rm A} &=&\frac{1}{m_{\rm A}\left(-\Omega^2+\Omega_0^2\right)},
\ea
and $c$ as the speed of light.
We have also included $\xi$, the motion of mirrors in the ring cavity, that are not due to quantum radiation-pressure noise (e.g. seismic or thermal noise).

Moreover, for the beam propagating through the amplifier in the reverse direction (i.e., from $a_i$ to $a_{{\rm IFO,}i}$), there is no amplification because the beam and pump counter-propagate. Hence, the input-output relation is a trivial phase shift~\eqref{eq:backward}. We defer a more complete discussion of the counter-propagating mode for \Cref{sec:counterProp}.

The MZ topology overcomes several challenges associated with optomechanical amplification, such as pump noise rejection (see~\Cref{sec:RIN}). One further advantage of the MZ configuration is that the pump and signal exit at separate ports. The signal can then be read out (e.g. using homodyne detection) without the strong amplifier pump field saturating the detection photodiodes.


\begin{figure}
    \includegraphics[width=\columnwidth]{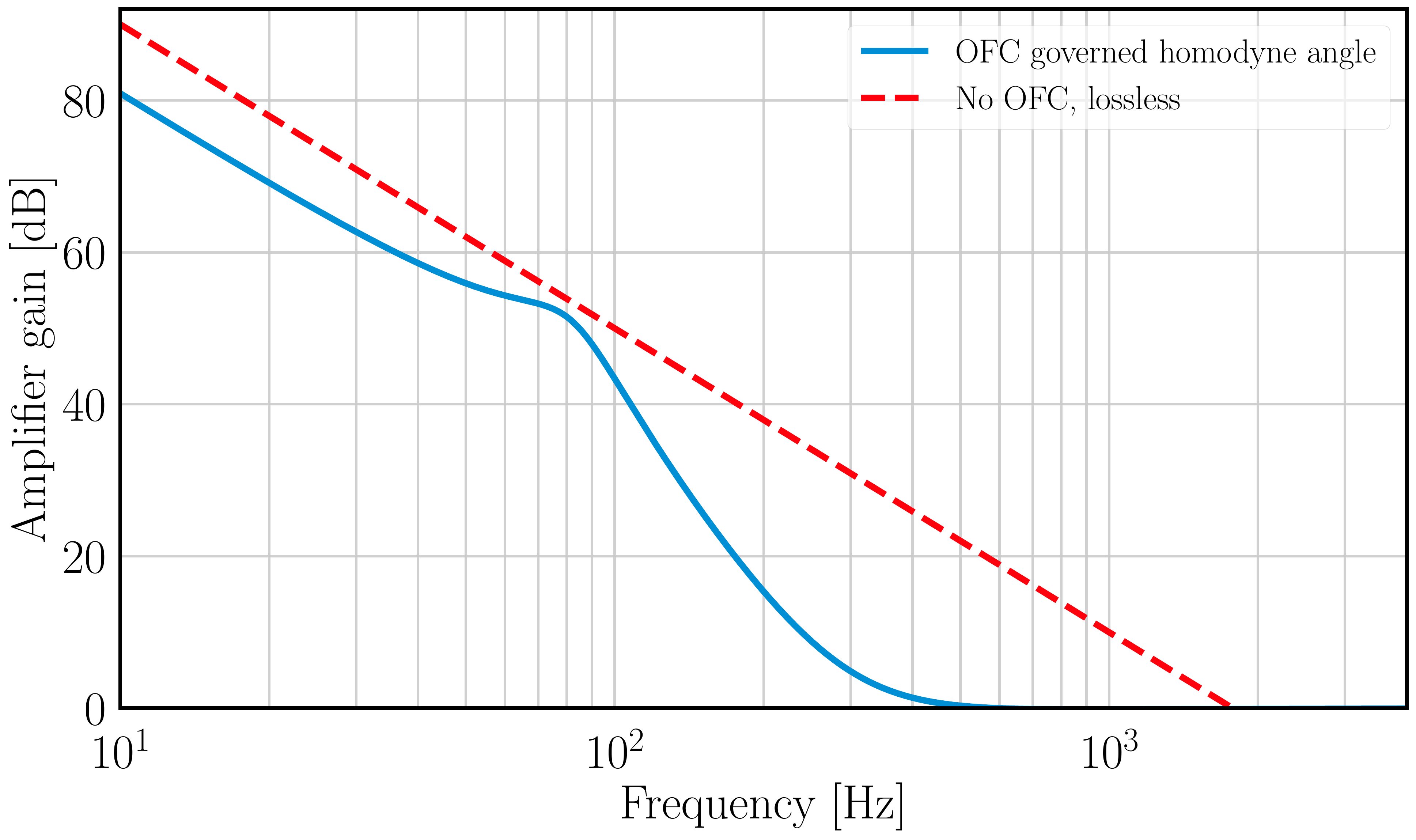}
    \caption{Optomechanical gain of the amplifier as a function of signal frequency, as seen in the frequency-dependent readout quadrature. For comparison, we have plotted the gain for both the simple case of a lossless amplifier with frequency independent readout quadrature (\Cref{eq:gain}), and a lossy amplifier with frequency dependent readout quadrature.} 
    \label{fig:amp_gain}
\end{figure}

We quantify the amplifier gain by considering the limit of a low-mass mirror and a strong pump, in which case the gain is simply the magnitude of the transfer function $\mathcal{K}_{\rm A}$ in~\eqref{eq:forward_approx}. Furthermore, we take the limit where the signal frequency is high compared to natural frequencies associated with the suspended mirror (typically $\lesssim 10\, {\rm Hz}$) so that the mechanical susceptibility is of order $|\chi_{\rm A}|\simeq (m_{\rm A}\Omega^2)^{-1}$, and we assume a wide cavity bandwidth $\Omega/\gamma_{\rm A}\ll 1$ so that
\begin{equation}
\label{eq:gain}
\left|\mathcal{K}_{\rm A}\right|\simeq
\left(\frac{0.01}{T_{\rm A}}\right)
\left(
\frac{30\,{\rm g}}{m_{\rm A}}
\right)
\left(
\frac{P_{\rm circ}}{40\,{\rm kW}}
\right)
\left(
\frac{1.5\,{\rm kHz}}{f}
\right)^2,\quad
\end{equation}
where $f=\Omega/2\pi$ is the signal frequency, and we assume carrier wavelength $\lambda_0=2\pi c/\omega_0=2\,{\rm \mu m}$. We find that the gain scales as $\propto 1/f^2$ with unity gain at $f\simeq 1.5\,{\rm kHz}$ for the characteristic $T_{\rm A}, \, m_{\rm A}, \text{and } P_{\rm source}\,$ given in~\eqref{eq:gain}. As expected, we achieve high gain at low frequencies $f\lesssim 500\,{\rm Hz}$ due to the high mechanical susceptibility of a low-mass mirror.

The power $P_{\rm circ}$ circulating in each ring is related to the source power $P_{\rm source}$ by
\ba\label{eq:I_circ_approx}
P_{\rm circ}&=&\frac{2}{T_{\rm A}}P_{\rm source}=40\,{\rm kW}\left(\frac{0.01}{T_{\rm A}}\right)\left(\frac{P_{\rm source}}{200\,{\rm W}}\right),
\ea
which is derived in~\Cref{app:ring} at~\eqref{eq:I_circ_exact}. Such a high source power can be achieved using a laser source of modest power by employing a power-recycling scheme (not shown in~\Cref{fig:MZ}).

\subsection{Synthesizing signal generation, amplification and detection}

As {\it input} to the amplifier, we assume that the GW signal is contained in only one quadrature, $b_{{\rm IFO },\, 1}$, which is the case for the Resonant Sideband Extraction (RSE) configuration in which Voyager is planned to be operated in. More specifically, we can re-write Equation (2.20) of \cite{BnC:2001} to get (in absence of optical losses)
\begin{align}
\label{eq:ifoio}
\left(
\begin{array}{c}
b_{\rm IFO,\,1} \\
b_{\rm IFO,\,2}
\end{array}
\right)
&=
e^{2i\Phi_{\rm IFO}}
\left(\begin{array}{cc}
1 & -\mathcal{K}_{\rm IFO} \\
0& 1
\end{array}
\right)
\left(\begin{array}{c}
a_{\rm IFO,\,1} \\
a_{\rm IFO,\,2}
\end{array}
\right) \nonumber\\
&+\sqrt{2\mathcal{K}_{\rm IFO}} e^{i\Phi_{\rm IFO}}
\left(\begin{array}{c}
1 \\
0
\end{array} \right)
\frac{h}{h_{\rm SQL}}
\end{align}
Losses in the interferometer were treated in detail in \cite{BnC:2001}, and will be further discussed in~\Cref{sec:IFO_noise}. Here, up to leading order in power loss $\mathcal{L}_{\rm IFO}$, we introduce additional fluctuations into the out-going field quadratures, in the same way as Eq.~(101) in Ref.~\cite{KLMTV:2001}, yielding
\begin{equation}
\label{eq:addifoloss}
b_{{\rm IFO},\,j} \rightarrow b_{{\rm IFO},\,j} + \sqrt{\mathcal{L}_{\rm IFO}} n_{{\rm IFO},\,j},
\end{equation}
where $n_{{\rm IFO},j}$ are vacuum fluctuations.

We further assume that frequency-dependent input squeezing of the $a_{\rm IFO,\,i}$ quadratures is applied, with the help of an Input Filter Cavity (IFC), such that quantum fluctuations in the signal quadrature $b_{\rm IFO,\,1}$ is suppressed at all frequencies. Note that we do not alter the frequency dependence of the input squeezed vacuum relative to the baseline Voyager design. Our amplifier is then designed to amplify $b_{\rm IFO,\,1}$.  More specifically, one can write
\begin{align}
    \left(\begin{array}{c}
a_{\rm IFO,\,1} \\
a_{\rm IFO,\,2}
\end{array}
\right) =&  e^{i\phi_{\rm IFC}}\mathcal{R}(\theta_{\rm IFC})     \left(\begin{array}{c}
e^{-2r } a_{\rm in,\,1} +\sqrt{\mathcal{L}_{\rm inj}} n_{\rm in,1} \\
e^{+2r } a_{\rm in,\,2}+\sqrt{\mathcal{L}_{\rm inj}} n_{\rm in,2}
\end{array}
\right)\nonumber\\
&+ \sqrt{\mathcal{L}_{\rm IFC}}
   \left(\begin{array}{c}
n_{\rm IFC,\,1} \\
n_{\rm IFC,\,2}
\end{array}
\right)
\end{align}
Here $\mathcal{R}$ is a rotation matrix
\begin{equation}
    \mathcal{R}(\theta)  =\left(
    \begin{array}{cc}
    \cos\theta & -\sin\theta \\
    \sin\theta & \cos\theta \end{array}
    \right)\,,
\end{equation}
 $e^{-2r}$ is the squeezing factor, $\mathcal{L}_{\rm inj}$ the injection loss of squeezing, and the rotation angle $\theta_{\rm IFC} =\arctan(\mathcal{K}_{\rm IFO})$ makes sure that the combined effect of $\mathcal{R}$ and the ponderomotive squeezing matrix in Eq.~\eqref{eq:ifoio} is to squeeze the $b_{\rm IFO,\,1}$ quadrature.  The loss $\mathcal{L}_{\rm IFC}$ is the loss of the IFC, introducing vacuum fluctuations, $n_{{\rm IFC},\,j}$.
 The optical parameters for the IFC that achieve the desired $15\,\rm{dB}$ of squeezing for Voyager are given in \Cref{tab:params}

One important difference between the  MZ amplifier and the original Caves proposal is that the quadrature it amplifies is frequency dependent; in other words, the anti-squeezing angle in the $b_{1,\, 2}$ quadrature basis is frequency-dependent.  In fact, the angle varies by as much as $90^\circ$ over the signal bandwidth.  At lower frequencies, $\mathcal{K}_A \gg 1$, and the $b_{{\rm IFO},\,1}$ quadrature is highly amplified, making the dominant contribution to $b_2$ (the $b_{{\rm IFO},\,2}$ contribution is comparatively negligible).  In this regime, the Caves proposal can be realized by detecting $b_2$.

However, by naively measuring the output quadrature $b_2$ at all frequencies, the signal is actually attenuated at high frequencies where $|\mathcal{K}_{\rm A}|\ll 1$, which is undesirable. Fortunately, this can be corrected by employing a frequency-dependent readout,
\begin{equation}
b_{\rm out} =b_\zeta=b_1\cos\zeta+b_2\sin\zeta
\end{equation}
which measures the amplified $b_2$ quadrature $(\zeta=\pi/2)$ at low frequencies and smoothly transitions to the unamplified $b_1$ quadrature $(\zeta=0)$ at high frequencies. This can be achieved by installing, between the amplifier and the readout chain, a filter cavity which we call the output filter cavity (OFC) (see~\Cref{fig:MZ}). The frequency-dependent reflection coefficient of the filter cavity, as well as its effect in rotating quadratures, is given in \Cref{eq:OFCrefl}; a more detailed treatment is given in Sec.~\G{IV C} of Ref.~\cite{KLMTV:2001}.   We expect that a $\simeq 40\,{\rm m}$ scale OFC is needed, with more detailed design parameters given in \Cref{tab:params}. Losses in the OFC also introduces additional vacuum fluctuations, which can be combined with the effect of the loss of readout photodetectors, leading to
\begin{equation}
\label{eq:bout}
    b_{\rm out} \rightarrow b_{\rm out} +\sqrt{\mathcal{L}_{\rm OFC}+\mathcal{L}_{\rm det}} n_{{\rm det},\,\zeta}
\end{equation}
Assuming that the noise that enters through this channel, $n_{{\rm det},\,\zeta}$, is unsqueezed vacuum, we plot the OFC-filtered amplifier gain in \Cref{fig:amp_gain}, showing no attenuation at high frequencies. 

To summarize: putting together Eq.~\eqref{eq:ifoio}--\eqref{eq:bout}, and inserting the amplifier input-output relation~\eqref{eq:forward_approx}, we can obtain the noise spectrum of the entire configuration, including losses from squeezing injection ($\mathcal{L}_{\rm inj}$), the IFC  ($\mathcal{L}_{\rm IFC}$), the interferometer ($\mathcal{L}_{\rm IFO}$, see Ref.~\cite{BnC:2001} and \Cref{sec:IFO_noise} for further details), the OFC ($\mathcal{L}_{\rm OFC}$), and the photodetectors ($\mathcal{L}_{\rm PD}$).  We will devote the next section, \Cref{sec:noise}, to noise in the amplifiers; there we will further introduce additional noise terms into Eq.~\eqref{eq:forward_approx}. 

\begin{figure}
    \centering
    \includegraphics[width=\columnwidth]{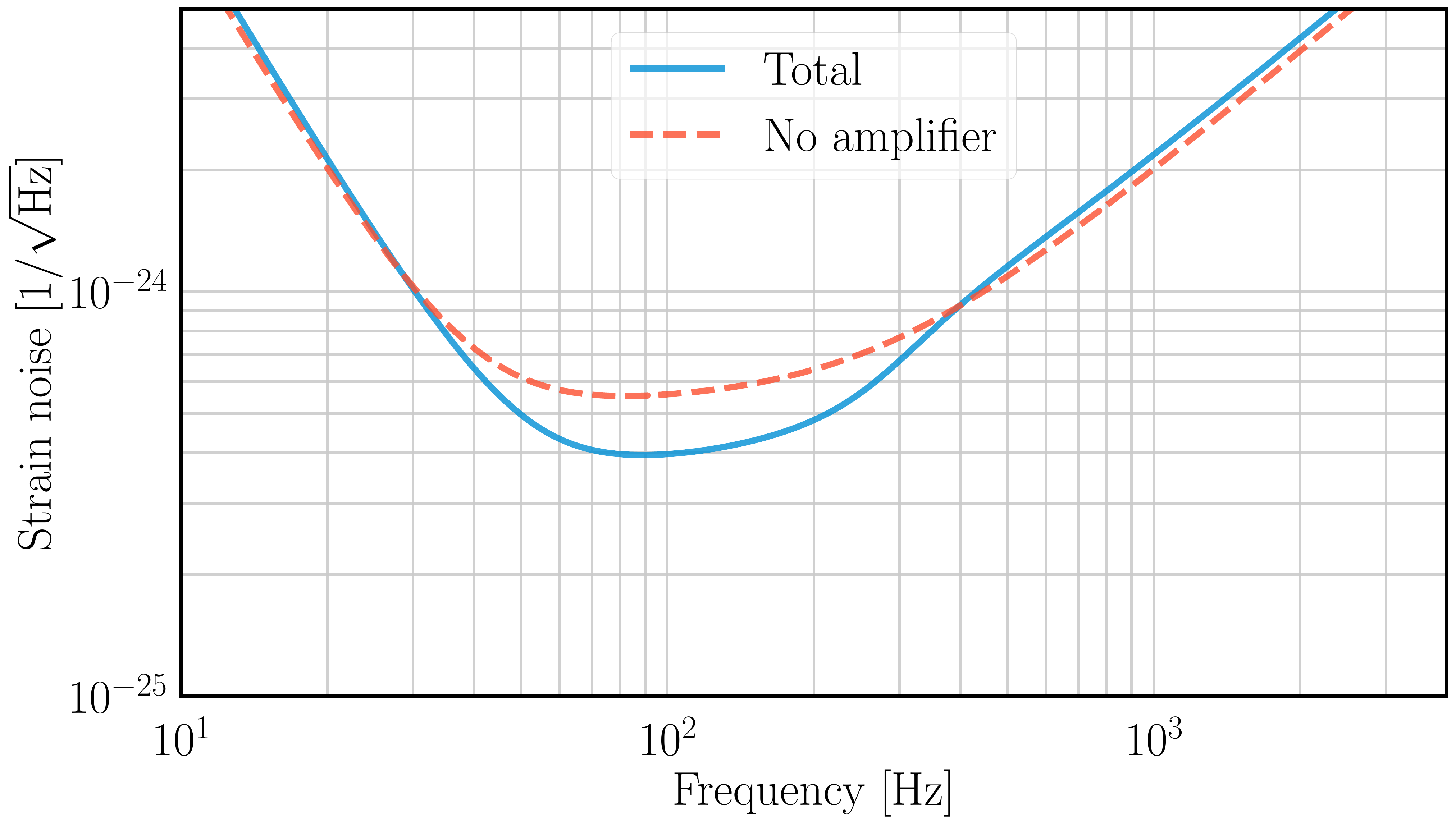}
    \caption{
    Comparison of Voyager sensitivity with (solid curve) and without (dashed curve) the amplifier. In both cases we assume 15 dB frequency-dependent squeezed vacuum injection. The design parameters are given in~\Cref{tab:params}. Sub-budgets for noise contributions from the amplifier and IFO may be found in~\Cref{fig:amp_budget} and~\Cref{fig:ifoNoises} respectively.
    }
    \label{fig:total_budget}
\end{figure}

Following the above steps, and incorporating discussions from the next two sections, we propose in~\Cref{tab:params} a set of design parameters for the amplifier, which were obtained by optimizing the total (amplifier \& IFO) noise using a cost function emphasizing the mid-band region $50\,{\rm Hz}\lesssim f \lesssim 500\,{\rm Hz}$, where the amplifier is most effective. The impact on the overall detector sensitivity is plotted in~\Cref{fig:total_budget}, showing modest improvement in this range. Given uncertainties in the parameters, our amplification strategy will be appropriate for application only if the interferometer's internal losses, test-mass coating thermal noise, achievable injected squeezed vacuum, and all other pre-amplification noises in Voyager turn out to be the same as or better than what we have chosen here, and if the readout photodetector inefficiency (and other losses from elements downstream of the amplifier, such as an Output Mode Cleaner) turn out to be the same as or worse than what we have chosen here.

Finally, we point out that our calculations remain valid only in the limit where the \emph{amplified} signal is much weaker than the pump, as expected since the pump is the source of the amplifier's energy. Since even for the strongest sources the signal \textit{power} is expected to be much weaker than the amplifier pump, we are well within the range of validity.

\section{Amplifier noise sources}
\label{sec:noise}

\begin{figure}
    \centering
    \includegraphics[width=\columnwidth]{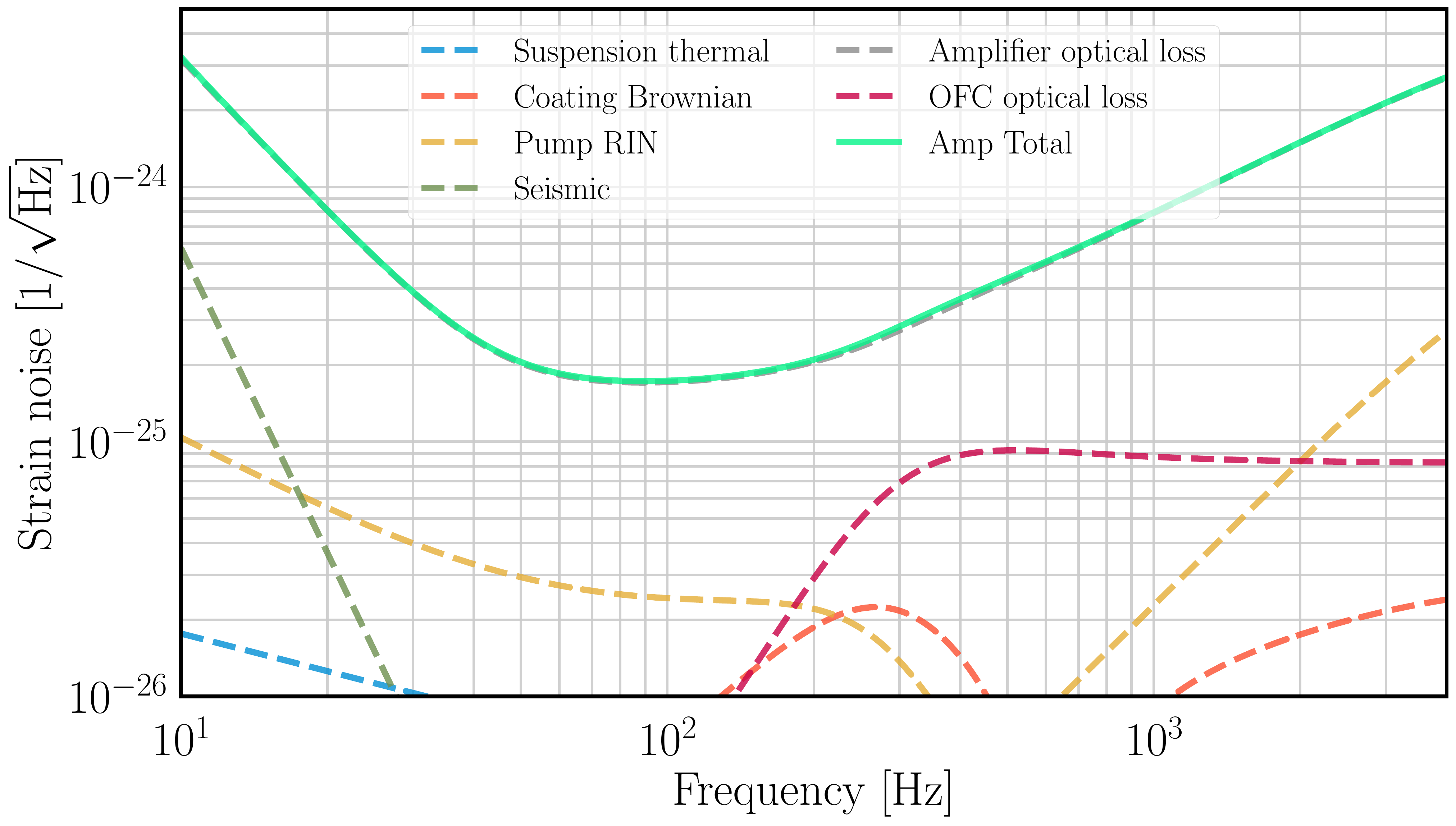}
    \caption{Breakdown of the dominant noise sources in the amplifier,
      including the OFC, based on parameters given in~\Cref{tab:params}
      for 15 dB frequency-dependent squeezed vacuum injection. The noise
      spectra as shown are projected into the GW strain sensitivity of
      the main interferometer.
      See~\Cref{sec:noise} for a detailed discussion of the various
      noise curves. The noise is dominated by optical loss in the
      ring cavities.}
    \label{fig:amp_budget}
\end{figure}

\begin{figure}
    \centering
    \includegraphics[width=\columnwidth]{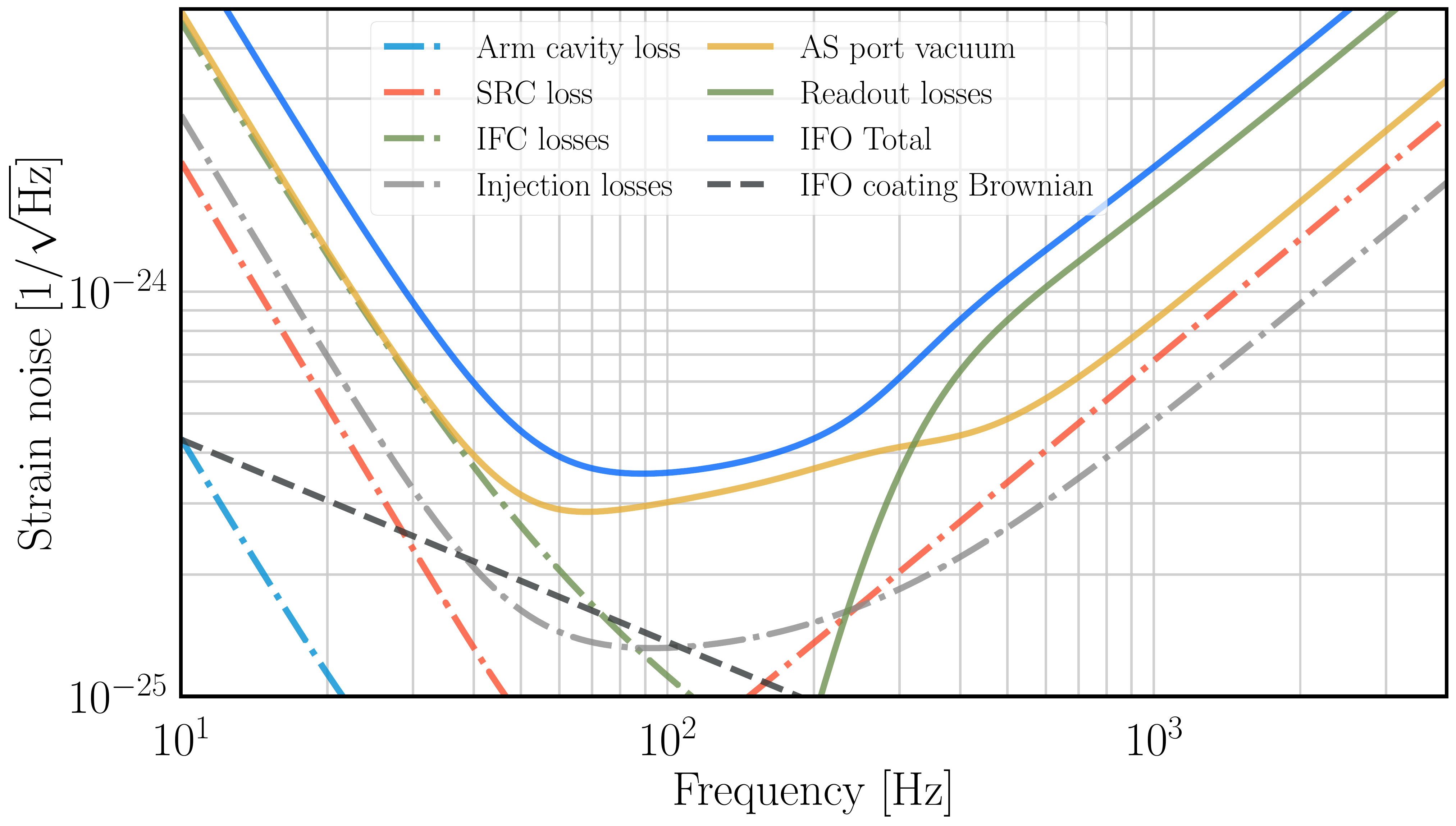}
    \caption{Breakdown of the dominant noise sources in the IFO, assuming the parameters given in~\Cref{tab:params} for 15 dB frequency-dependent squeezed vacuum injection. The noise levels are signal-referred assuming the amplifier is installed. See~\Cref{sec:IFO_noise} for a detailed discussion of the various noise sources. }
    \label{fig:ifoNoises}
\end{figure}

In order for the amplifier to be effective, it must not introduce significant noise sources of its own. In this section, we discuss some of the dominant amplifier noise sources, such as optical losses in the ring cavities, pump intensity fluctuations, backscattering of pump off amplifier optics, and finally the coating Brownian noise and suspension thermal noise of the amplifier optics. The associated noise curves are given in~\Cref{fig:amp_budget}, assuming the parameters given in~\Cref{tab:params} for 15 dB squeezed vacuum injection. We find, under these assumptions, that the amplifier noise is dominated by optical loss in the ring cavities.


\subsection{Optical losses in the ring cavities}
\label{sec:Losses}
As the signal beam circulates within the the ring cavities, dissipative and transmissive losses are accumulated at each optic. The small amount of unsqueezed vacuum that enters the signal mode in this process is amplified optomechanically, thus causing the squeezed vacuum to decohere. It is therefore crucial to keep the total loss as low as possible.

Transmissive loss through each high-reflective (HR) mirror, labelled M2R, M2L, M3R \& M3L in~\Cref{fig:MZ}, is limited to 5 ppm by a suitable dielectric bi-layer coating. Dissipative losses arise due to a number of mechanisms, but may broadly be grouped into absorption or scatter. We assume absorption in the amplifier mirror substrate and dielectric coatings will be $\lesssim 1\, \mathrm{ppm}$ per optic.

Modeling and characterizing loss due to scattering is an area of active research. An empirical scaling law has been found to describe the measured round-trip scatter loss in a variety of two-mirror optical cavities~\cite{Isogai2013, Isogai:2016}, given by
\ba
\label{eq:scatterLoss}
\frac{\mathscr{L}}{\mathrm{ppm}} = \left (\frac{4 \pi}{\lambda/\mathrm{nm}} \right ) ^2 \frac{A}{\mathrm{nm}^2 \cdot \mathrm{mm}}  \frac{1}{\gamma {-}1} \left ( \frac{\mathrm{mm}}{\sqrt{2}\alpha w}  \right) ^{1 - \gamma},
\ea
where $w$ denotes the Gaussian beam radius, while  $A = 8 \times 10^{-3}\, \mathrm{nm}^2 \cdot \mathrm{mm}$ and $\gamma = 1.2$ are model parameters extracted from measurements. They parametrize the power spectral density (PSD) of micro-roughness on the mirror surface, which is assumed to obey $\mathrm{PSD}(f_s) = A \left( \frac{f_s}{1/\mathrm{mm}} \right ) ^{-\gamma}$ for spatial frequencies $f_s$ greater than some cutoff. This cutoff frequency is modelled as $f_s^{\mathrm{min}} = \frac{1}{\alpha w}$, with $\alpha=2$ corresponding to the Gaussian beam diameter. The contribution to scatter from spatial frequencies smaller than the cutoff are neglected. With $\lambda=2\,\mu{\mathrm{m}}$, $w = 5\, \mathrm{mm}$ and $\alpha=1$ in~\eqref{eq:scatterLoss}, we estimate $\simeq 3\,\mathrm{ppm}$ of scatter loss per optic. Since understanding of the impact of scattered light is evolving~\cite{Zeidler:17}, we choose for our analysis the round number $5\,\mathrm{ppm}$ of scatter loss per optic.

In summary, we assume a \textit{total} of 30\,ppm loss per round-trip in each ring cavity, accounting for the various mechanisms described in this section.
This loss introduces additional vacuum fluctuations to $b_{1,2}$ in Eq.~\eqref{eq:forward_approx}, in the same way as 
Eq.~\eqref{eq:addifoloss}.

\subsection{Pump intensity noise}
\label{sec:RIN}



Power fluctuations of the amplifier pump laser produce radiation pressure fluctuations on the ring cavity mirrors, and are thereby amplified by the optomechanical feedback explained in~\Cref{sec:amp}. With perfectly symmetric ring cavities, the MZ topology has the advantage of separating the pump noise, which couples only to the field labelled $n_i$ in~\Cref{fig:MZ}, from the signal, which couples only to the field labelled $b_i$ in~\Cref{fig:MZ}. In practice, however, slight differences between the two ring cavities lead to imperfect common-mode rejection.
This noise can be estimated by first computing the spectrum of mirror of motion due to radiation-pressure fluctuations, propagating it through Eq.~\eqref{eq:forward_approx} (i.e., add this displacement noise contribution to the noise spectrum of $\xi$), and then suppressing it by a common mode rejection factor.

For our simulations, we crudely model the amplitude spectral density of the relative intensity noise (RIN) of the amplifier pump according to
\ba
\label{eq:RIN}
\mathrm{RIN}(f) = \left | \frac{f + f_0}{f} \right | \frac{1 \times 10^{-9}}{\sqrt{\mathrm{Hz}}}, \quad f_0 = 50\,\mathrm{Hz}.
\ea
Meeting this requirement on laser intensity noise is expected to be challenging. In order for the relative intensity noise due to shot noise on a sensing photodiode at $2\,\mu \mathrm{m}$ to be $\lesssim 1 \times 10^{-9}\,{\rm Hz}^{-1/2}$, we would need to detect $\simeq 300\,\mathrm{mW}$ of power on that photodiode, corresponding to a dynamic range of $\gtrsim 10^9$. Nevertheless, promising techniques have been demonstrated~\cite{Henning2018, Kwee2009}, and we anticipate that sufficient progress will be made to achieve this level of stabilization.

Furthermore, we assume 60\,dB common-mode rejection. In our simulations, the asymmetry is modeled as a difference in the ring cavity finesse. Realizing this level of common-mode noise rejection is challenging, but has been achieved in terrestrial gravitational wave detectors.
Moreover, we expect to be able to tune the finesse of each cavity by $\simeq 1\,\%$, for instance by changing the spot positions on the cavity mirrors to sample regions of slightly different optical loss. 



\subsection{Noise from the counter-propagating mode}
\label{sec:counterProp}
As discussed in~\Cref{sec:Losses}, surface roughness and point defects on the amplifier optics can scatter the high power circulating pump field out of the resonant cavity mode. Some portion of this scattered light then becomes resonant in the \textit{counter-propagating} mode of the ring cavities. Subsequently, this field leaves the amplifier and is injected directly back into the main interferometer via the anti-symmetric port, where it mixes with the squeezed vacuum quadratures $a_{{\rm IFO},\,i}$ (see~\Cref{fig:MZ}). This gives rise to noise in the readout due to two effects: (1) displacement noise of the amplifier optics, and (2) amplifier pump noise. We expect that the amplifier optics are sufficiently well isolated from displacement noise, and focus instead on the latter with emphasis on the relative intensity noise discussed in~\Cref{sec:RIN} in the context of common-mode rejection.

Measurements on the Advanced LIGO output mode cleaner cavity, which has an angle of incidence of approximately 4 degrees, suggest that less than 1\,ppm of the incident field is retro-reflected~\cite{Martynov:2015}. Since the ring cavities in our design have a 30 degree angle of incidence at each optic, the back-scatter is expected to be smaller.  For our calculations, we assume a fraction $\mathcal{E}_{\rm bs}=10^{-7}$ of the pump power is back-scattered into the anti-symmetric port. This field then adds noise to the input quadratures $a_{{\rm IFO},\,i}$ which we can estimate by
\ba\label{eq:bs}
\sqrt{S_{a_{{\rm IFO},i}}^{\mathrm{(bs)}}}&=&\sqrt{\frac{1}{2}\mathcal{E}_{\rm bs}\left(\frac{P_{\rm source}}{2\hbar\omega_0}\right)
}
\mathrm{RIN}(f)
,
\ea
where the overall factor $1/\sqrt{2}$ splits the noise evenly into the two quadratures. With $P_{\rm source}=200\,{\rm W}$ and ${\rm RIN}=10^{-9}/\sqrt{\rm Hz}$, we find that~\eqref{eq:bs} evaluates to $\simeq 7\times 10^{-3}$ per quadrature, which is negligible even for the case of 20 dB ($=10^{-1}$) squeezed vacuum injection. Hence, it is omitted from our analysis.

There are other possible mechanisms of scattered light degrading the interferometer sensitivity, particularly given the proximity of the high power amplifier pump field to the interferometer's anti-symmetric port. For instance, some of the scattered light could leave the ring cavity, scatter off the vacuum chamber walls, and recombine into the cavity's signal mode. Acoustic and seismic vibrations of the walls then lead to phase modulation of the back-scattered light. Problems of this nature may be addressed by installing baffles on the walls, as was done for the LIGO beam tubes. Another possible noise coupling mechanism is due to intensity fluctuations on the resonant counter-propagating mode displacing the amplifier mirrors via radiation-pressure. We find that the phase noise thus induced is below the level of the seismic noise for the amplifier systems considered in this paper, but this could become a significant noise source for amplifiers that use much lighter mirrors. We leave the detailed analysis of such noise coupling mechanisms to future work.




\subsection{Coating Brownian noise}
\label{sec:coatBr}
Thermal fluctuations of the dielectric coatings~\cite{Hong:2013} on the amplifier optics produce phase fluctuations on the reflected beam, which manifests as noise in the amplifier readout.
Mathematically, this contributes through $\xi$ in Eq.~\eqref{eq:forward_approx}.

To mitigate this effect, we propose for the high reflectivity mirrors (i.e., M2L, M3L, M2R \& M3R in~\Cref{fig:MZ}) a coating structure comprising of alternating layers of silicon nitride (SiN) and amorphous silicon (aSi). This choice was motivated by the promising mechanical loss of SiN at the proposed operating temperature of 123\,K~\cite{Steinlechner:2018, Southworth:2009}.
Nevertheless, more work is needed to determine its feasibility and, in particular, to lower losses due to absorption in the SiN layers. Rather than using the canonical quarter wave stack to realize the HR coating, a numerical optimization algorithm was used to identify the thickness of each layer in a 12 bi-layer pair stack in order to optimize the resulting noise~\cite{coatingPaper:2019}. For the assumed mechanical properties of aSi/SiN (see \Cref{tab:params}), and meeting the coating power transmissivity requirement of $\lesssim5\,\mathrm{ppm}$, we estimate a Brownian noise contribution per optic that is $\simeq 3.5\,{\rm dB}$ below that of an $\alpha$-Si/SiO$_2$ coating with the same power transmissivity. Note that we neglect the Brownian noise contribution from the M1R \& M1L optics in~\Cref{fig:ring}, since they have higher transmissivity and therefore require fewer dielectric layers.

While coating Brownian noise in the amplifier ring cavities is not the dominant noise source, we find that in order to take full advantage of the sensitivity improvement offered by the amplifier, the coating Brownian noise of the test masses will have to be improved by a factor of $\simeq$4\,--\,5 from the current  design. This is further discussed in \Cref{sec:IFO_coating}.




\subsection{Suspension thermal noise}
\label{sec:suspTherm}

Similar to coating Brownian noise,
any fluctuations of the position of the mirrors in the amplifier, also enter the output of the amplifier through $\xi$ in Eq.~\eqref{eq:forward_approx}.

To isolate the amplifier mirrors from seismic vibrations, we propose a double pendulum suspension for each optic in the ring cavities. Internal friction in the fibers couple environmental thermal fluctuations to mirror displacement. For this analysis, we consider only the thermal noise due to the lower (of the double-stage) suspension fiber. The key parameter characterizing internal friction in the fibers is the frequency-dependent loss angle $\phi(\omega)$ which has contributions from (a) the surface, (b) the bulk and (c) thermoelastic effects~\cite{Cumming2013}. For high-purity silicon fibers at 123\,K, the thermoelastic and bulk loss contributions are negligible. The loss angle is then dominated by surface imperfections and defects, which can be modeled by a characteristic depth $h$ and a surface loss angle $\phi_s$. We assume $h=1\,\mu \mathrm{m}$ and $\phi_s = 10^{-5}$, and follow the formalism described in~\cite{Cumming2013} to evaluate the loss angle. Finally, the amplitude spectral density of this displacement noise can be obtained by applying the fluctuation-dissipation theorem~\cite{PhysRev.83.34}, as
\ba
\label{eq:suspThermNoise}
    x(\omega) = \sqrt{\frac{4 k_B T}{\omega m_{\rm A} } \left ( \frac{\omega_0^2 \phi(\omega)}{\omega_0^4 \phi^2 (\omega) + [\omega_0^2 - \omega^2]^2} \right )},
\ea
where $T$ is the equilibrium temperature of the system, and $f_0=\omega_0/2\pi$ is the resonant frequency of the suspension. A more thorough analysis requires Finite Element Analysis (FEA) calculations to validate the analytic approximations. In our present modeling, we have a large safety factor for this noise contribution, and so are immune to it being higher by a factor of a few (the analytic calculation is expected to be accurate to within this factor).



\begin{table}[]
    \begin{tabular}{ c|lr }
    \hline
    \hline
     & Parameter & Value \\
    \hline
    \multirow{4}{*}{\rotatebox[origin=c]{90}{IFO}} & Arm cavity round-trip loss & 20 ppm \\
     & SRC round-trip loss & 300 (100) ppm \\
     & Readout chain loss & 10 \% \\
     & Carrier wavelength & $2 \,\mu \mathrm{m}$ \\
    \hline
    \multirow{6}{*}{\rotatebox[origin=c]{90}{SQZ}} & Squeeze injection & 15 (20) dB \\
     & Injection loss & 1 (0.3) \% \\
     & IFC round-trip loss  & 20 (10, 10) ppm \\
     & IFC length  & 500 (800, 800) m \\
     & IFC detuning & -33.4 (-34.6, 4.96) Hz \\
     & IFC input coupler transmission & 0.14 (0.22, 0.22) \% \\
    \hline
    \multirow{11}{*}{\rotatebox[origin=c]{90}{AMP}} & Ring cavity round-trip loss, $\mathscr{L}$ & 30 (15) ppm \\
     & M1L/M1R transmissivity, $T_{\rm A}$ & 0.89 (0.90) \% \\
     & Round-trip cavity length, $L_{\rm A}$ & 30 m \\
     & Pump source power, $P_{\rm source}$ & 220 (230) W \\
     & Mirror mass, $m_{\rm A}$ & 30 (10) g \\
     & Mirror substrate & Si \\
     & Common mode rejection & 60 dB \\
     & OFC round-trip loss & 20 (10) ppm \\
     & OFC length & 40 (25) m \\
     & OFC detuning & -80.4 (-77.8) Hz \\
     & OFC input coupler transmission & 43 (22) ppm \\

    \hline
    \multirow{6}{*}{\rotatebox[origin=c]{90}{COAT}} & Refractive index of aSi, $n_{H}$ & 3.65 \cite{Steinlechner:2018}  \\
     & Reflactive index of SiN, $n_{L}$ & 2.17 \cite{Steinlechner:2018} \\
     & Number of layer pairs & 12 \\
     & Mechanical loss of aSi (123 K) & $3 \times 10^{-5}$ \cite{Steinlechner:2018} \\
     & Mechanical loss of SiN (123 K) & $2 \times 10^{-5}$ \cite{Steinlechner:2018} \\
     & Beam radius, $w_{\mathrm{beam}}$ & 5 mm \\
     \hline
    \multirow{15}{*}{\rotatebox[origin=c]{90}{SUS}} & Material & Silicon  \\
    & Width & 250 $\mu \mathrm{m}$ \\
    & Thickness & 50 $\mu \mathrm{m}$ \\
    & Number of fibers & 2 \\
    & Length of pendulum & 60 cm \\
    & Surface loss angle, $\phi_s$ (123 K) & $10^{-5}$ \\
    & Bulk loss angle, $\phi_{\mathrm{bulk}}$ (123 K) & $2 \times 10^{-9}$ \\
    & Surface depth, $h$ & 1 $\mu \mathrm{m}$ \\
    & Young's modulus & 155.8 GPa \\
    & Coefficient of thermal expansion & $10^{-10} \mathrm{K}^{-1}$ \\
    & ${d \log Y}/{dT}$ & $-2 \times 10^{-5} \mathrm{K}^{-1}$ \\
    & Heat capacity & 300 $\mathrm{J} \ \mathrm{kg}^{-1} \  \mathrm{K}^{-1}$\\
    & Thermal conductivity & 700 $\mathrm{W} \ \mathrm{m}^{-1} \  \mathrm{K}^{-1}$ \\
    & Density & 2329 $\mathrm{kg} \ \mathrm{m}^{-3}$ \\
    \hline
    \hline
    \end{tabular}
    \caption{Summary of design parameters used in our simulations for the main interferometer (IFO), squeezed vacuum injection path (SQZ), amplifier optomechanics (AMP), amplifier HR coating (COAT) and amplifier suspension (SUS), assuming 15 dB squeezed injection. Design parameters for 20 dB (see App.~\Cref{app:20dB}) are also included in parentheses wherever they differ from 15 dB. Note that for 20 dB, two IFCs are required and so the two comma-separated numbers refer to the first and second (as seen from the squeezed vacuum source) IFCs respectively. Many of the material properties in the COAT and SUS subsystems are still under study, and so may turn out to be significantly different from what we have assumed.
    }
    \label{tab:params}
\end{table}

\section{Noise in the main interferometer}
\label{sec:IFO_noise}

In order for the amplifier to be effective, the detector sensitivity must be limited by optical losses in the readout chain. In this section, we discuss some noise sources in the IFO, including a variety of other optical losses and coating Brownian noise of the test masses, which in the current Voyager design are at a level comparable to readout losses.
A plot of these noise curves is given in~\Cref{fig:ifoNoises}, assuming the parameters in~\Cref{tab:params} for 15\,dB squeezed injection.

\subsection{Optical losses}
\label{sec:IFOloss}

In addition to readout loss, we consider a variety of optical losses within the main interferometer optics, consisting of four independent contributions:
(a) `injection losses' to be explained below,
(b) input filter cavity (IFC) losses,
(c) signal recycling cavity (SRC) losses, and
(d) arm cavity losses.

The `injection loss' lumps together the insertion loss of (i) two Faraday isolators (one double-pass and one single-pass) between the squeezed vacuum source and the interferometer, and (ii) mismatch between the SRC and IFC spatial modes. We assume a total injection loss of 1\,\%, anticipating that the insertion loss of the Faraday Isolators can be improved to 0.2\,\% per pass.

The SRC loss can be broken down further into
(i) the spatial mode mismatch between the SRC and the interferometer's differential mode,
(ii) absorption in the IFO beamsplitter (BS) and corner mirror (ITM) substrates, and
(iii) reflection from the anti-reflective (AR) coatings on the BS and ITMs.
In particular, we require less than 300\,ppm total round-trip loss for the SRC. Assuming that substrate absorption in, and reflection from the AR coatings for the ITMs and BS add up to a total loss of 50\,ppm, the requirement on the mode mismatch to the interferometer's differential mode is 250\,ppm, which is expected to be challenging considering that the lowest achieved mode mismatch in the current generation of interferometers is $\simeq 1\%$.

Our choices for the losses represent an optimistic estimate of the
progress in loss reduction over the next 10\,years. In particular, we
are emphasizing the situation in which a high level of squeezing is injected,
but the readout losses are high. In the case where injection losses are high and readout losses are low, the efficacy of the amplifier is substantially reduced (or even non-existent).
A summary of our design requirements for the losses is given in~\Cref{tab:params}, with the corresponding noise curves given in~\Cref{fig:ifoNoises}.

\subsection{Coating Brownian noise in Voyager test masses}
\label{sec:IFO_coating}


According to the current Voyager design, which only assumes 10\,dB of squeezed vacuum injection~\cite{Voyager:Inst}, it is anticipated that coating Brownian noise of the interferometer's arm cavity optics will be the dominant noise contribution from 40\,--\,100\,Hz.
This would impose a serious limit on the performance of all quantum non-demolition schemes, not just the amplifier proposed in this work. Furthermore, in our analysis, we have considered even higher levels of squeezed vacuum injection, and so we expect an even broader band over which coating Brownian noise will be dominant.
This is especially problematic since the amplifier is designed to be most effective in the same frequency band (see~\Cref{fig:total_budget}).
To fully take advantage of the sensitivity improvement offered by the amplifier, we have assumed that the coating Brownian noise can be reduced by a further factor of 4\,--\,5 at all frequencies.
This more speculative coating Brownian noise is plotted in~\Cref{fig:ifoNoises} (and~\Cref{fig:speculative} for 20\,dB, as discussed in~\Cref{app:20dB}).
While achieving this reduction is a challenging prospect, several promising leads are being explored~\cite{reid2016development}.

\section{Conclusions and Experimental Outlook}
\label{sec:conclusion}
In this work, we have presented a novel approach to amplify the GW signal to protect the signal to noise ratio against quantum decoherence. We emphasize that this work is only meant to be an exploratory study of the idea of using an optomechanical device as a phase sensitive amplifier for terrestrial GW detectors. In order for this to be incorporated into the design of a detector such as Voyager, significant work will have to be done, starting with table-top scale experiments, to get a deeper understanding of the many technical noise sources we have not addressed in this paper (e.g. feedback control of the amplifier, losses due to the mode matching between the interferometer and the amplifier, fluctuations of the relative alignment of the cavities, imperfections in the amplifier optics, etc.).

It has been previously been proposed to use nonlinear crystals to amplify the interferometer output~\cite{korobko2018engineering}, and to use atomic systems to generate ``negative inertia'' that leads to back-action evasion~\cite{khalili2018overcoming}. These are promising approaches, but requires mitigating (i) backscatter noise due to the lower optical quality of crystals relative to super-mirrors, (ii) high scatter loss induced decoherence due to the poor optical quality, and (iii) pump noise coupling due to a non-vacuum seed.

Promising future directions to consider include:
\begin{enumerate}
  \item lighter masses in the amplifier (where the optomechanical gain would extend to higher frequencies),

  \item interferometers (with higher masses) which are limited by shot noise rather than radiation pressure at lower frequencies (where there is already significant optomechanical gain),

    \item a hybrid diplexed crystal-optomechanical approach where a crystal amplifier is used to extend the optomechanical amplifier gain to higher frequencies.
\end{enumerate}

\section*{Acknowledgments}
We would like to acknowledge conversation with the Quantum Noise and Advanced Interferometer working groups of the LIGO Science Collaboration.
RXA thanks Carl Caves for several stimulating conversations about evading quantum mechanics.
Plots were produced with \texttt{matplotlib}~\cite{Hunter:2007}.

YB is supported by the Sherman Fairchild Fellowship of the Walter Burke Institute for Theoretical Physics and by the U.S. Department of Energy, Office of Science, Office of High Energy Physics, under Award Number DE-SC0011632.
YC is supported by NSF grants PHY-1612816, PHY-1708212 and PHY-1708213, and the Simons Foundation (Award Number 568762).
RXA, GV, CW, AM, and ARW were supported by PHY-0757058.
LIGO was constructed by the California Institute of Technology and Massachusetts Institute of Technology with funding from the National Science
Foundation, and operates under cooperative agreement PHY-0757058.
KK and RXA were supported by Boeing (Award Number CT-BA-GTA-1).

\appendix

\section{Derivation for the input-output relations
  of the Mach-Zehnder amplifier}
\label{app:ring}

\begin{figure}
    \centering
    \includegraphics[width=\columnwidth]{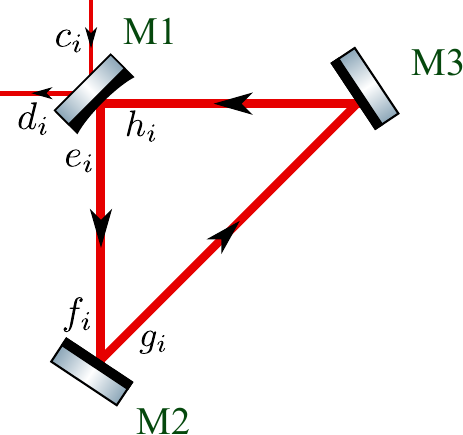}
    \caption{A three-mirror triangular ring cavity with pump and signal co-circulating counterclockwise (not drawn to scale). We assume all three mirrors have the same mass $m_{\rm A}$. Two of these ring cavities are combined to form the MZ amplifier in~\Cref{fig:MZ}}.
    \label{fig:ring}
\end{figure}

In this section, we provide a detailed derivation of the input-output relation for the MZ amplifier introduced in \Cref{sec:MZ}. While the effect of optical losses, mirror displacement noise, and other technical noise sources are neglected in the calculations presented in this section, they are included in our numerical simulations.

We begin by computing the input-output relation for one of the two triangular ring cavities whose field amplitudes are labeled in \Cref{fig:ring}. We consider the case where the pump and signal co-propagate counterclockwise. We let $c_{1,2}$ and $d_{1,2}$ denote the incident and reflected signal field amplitudes, respectively, and we assume the incident pump (electric) field has the form $E(t)=E_0\cos(\omega_0 t)$ for some constant amplitude $E_0$.  We let $r_{\rm A} = \sqrt{1 - T_{\rm A}}, \, t_{\rm A}=\sqrt{T_{\rm A}}$ denote the amplitude reflection and transmission coefficients of mirror M1, respectively, as seen from the cavity interior. The amplitude reflection coefficient of M2 and M3 is set to $+1$ (again as seen from the interior of the cavity). We let $l_1$ denote the distance that the beam travels from M1 to M2, and $l_2$ the distance that the beam travels from M2 to M3 and then to M1. Furthermore, we assume no cavity detuning. Under these conditions, the pump resonates in the cavity, thus enhancing the radiation pressure effect we are exploiting.

The beating between the pump and signal as they co-propagate along the ring produces the optomechanical amplification discussed in \Cref{sec:opto}. The total amplification is equivalent to that produced by a cavity where M1 and M3 are fixed and where the boundary condition at the free mirror M2 is
\ba\label{eq:M2_bc}
g_1=f_1,\quad g_2=
-\kappa_{\rm A} f_1+f_2,
\ea
with
\ba
\kappa_{\rm A}&=&-\frac{8\omega_0 P_{\rm circ}}{c^2}\sum_{i=1}^3\cos^2\left(
\theta_{{\rm inc}, i}\right)\chi_{i},
\ea
where $P_{\rm circ}$ is the power circulating in the ring, and for each optic M$_i$ ($i=1,2,3$) we have $\theta_{{\rm inc},i}$ as the angle of incidence and $\chi_{i}$ as the mechanical susceptibility. For a detailed derivation of~\eqref{eq:M2_bc} in the limit where the mirrors are free masses, see Sec. IV. A. of~\cite{Corbitt2005}. In our simulations, we assume that all three mirrors have the same mechanical susceptibility, $\chi_{\rm A}$, and that the ring is arranged as an equilateral triangle configuration (i.e., $\theta_{{\rm inc},i}=\pi/6$), in which case $\kappa_{\rm A}$ reduces to~\eqref{eq:ring_K}. Furthermore, the boundary conditions at M1 are given by
\ba
d_1&=&t_{\rm A} h_1-r_{\rm A} c_1,\quad d_2=t_{\rm A} h_2-r_{\rm A}c_2,\nonumber\\
e_1&=&t_{\rm A} c_1+r_{\rm A}h_1,\quad e_2=t_{\rm A} c_2+r_{\rm A}h_2,
\ea
and the field amplitudes propagate along the ring as
\ba
f_1&=&e_1e^{i\Omega l_1/c},\quad f_2=e_2e^{i\Omega l_1/c},\nonumber\\
h_1&=&g_1e^{i\Omega l_2/c},\quad h_2=g_2e^{i\Omega l_2/c}.
\ea

Solving this system gives us the input-output relation for the ring as
\ba\label{eq:ring_IO}
\begin{pmatrix}
d_1\\
d_2
\end{pmatrix}
=
e^{i2\eta}
\begin{pmatrix}
1 & 0 \\
-\mathcal{K}_{\rm A} & 1
\end{pmatrix}
\begin{pmatrix}
c_1\\
c_2
\end{pmatrix},
\ea
where
\ba
e^{i2\eta}&=&\frac{e^{i\Omega L_{\rm A}/c}-r_{\rm A}}{1-e^{i\Omega L_{\rm A}/c}r_{\rm A}},\nonumber\\
\mathcal{K}_{\rm A}&=&
\left(
\frac{
t_{\rm A}^2
}{
1
-2\cos(\Omega L_{\rm A}/c)r_{\rm A}
+r_{\rm A}^2
}
\right)
\kappa_{\rm A}
,
\ea
with $L_A=l_1+l_2$ as the round-trip length of the ring.

For the full MZ configuration of \Cref{fig:MZ}, we need to analyze the input-output relations for both the `forward' direction (i.e., from $b_{{\rm IFO,}i}$ to $b_i$) and the `backward' direction (i.e., from $a_i$ to $a_{{\rm IFO},i}$). The relations for the forward direction can be trivially derived from~\eqref{eq:ring_IO} to be
\ba\label{eq:forward}
\begin{pmatrix}
b_1\\
b_2
\end{pmatrix}
=
e^{i2\eta}
\begin{pmatrix}
1 & 0 \\
-\mathcal{K}_{\rm A} & 1
\end{pmatrix}
\begin{pmatrix}
b_{\rm IFO,1}\\
b_{\rm IFO,2}
\end{pmatrix}.
\ea
In the limit $\Omega L_A/c\ll1$ where the cavity is short compared to a signal wavelength, and in the limit $t_{\rm A}^2\ll 1$ where M1 is highly reflecting, the forward relation reduces to~\eqref{eq:forward_approx}, whereby the results are expressed in terms of the power reflectivity $R_{\rm A}=r_{\rm A}^2$ and power transmissivity $T_{\rm A}=t_{\rm A}^2$. For the backward direction, the beam is immune to the optomechanical effect since it counter-propagates against the pump, and so the relations are like \eqref{eq:forward} but with the pump `turned off', giving
\ba\label{eq:backward}
\begin{pmatrix}
a_{\rm IFO,1}\\
a_{\rm IFO,2}
\end{pmatrix}
=
e^{i2\eta}
\begin{pmatrix}
a_{1}\\
a_{2}
\end{pmatrix}.
\ea
While amplitude fluctuations of the pump induce mirror displacement fluctuations that couple to the phase of the counter-propagating mode, these fluctuations can be ignored since the counter-propagating mode is weak to begin with, and the mirror's displacement is small compared to a carrier wavelength. Finally, we must also deduce the relation between the power $P_{\rm circ}$ circulating in each ring and the source power $P_{\rm source}$ of the amplifier, as labeled in~\Cref{fig:MZ}. A simplified version of the above calculation gives
\ba\label{eq:I_circ_exact}
P_{\rm circ}&=&\frac{1}{2}\left(\frac{t_{\rm A}}{1-r_{\rm A}}\right)^2P_{\rm source},
\ea
where the overall factor $1/2$ takes into account the beamsplitting at BS1. In the limit $T_{\rm A}\ll 1$ this reduces to~\eqref{eq:I_circ_approx}.

\section{Output filter cavity reflectivity}

The reflectivity of a filter cavity has been derived in Appendix A of \cite{Evans:2013}, while a more general treatment of its quadrature rotation effects were given in Appendix A of \cite{purdue2002practical}.  We have adopted the equations derived there, but due to a difference in Fourier transform convention, have used the complex conjugate of Equation A6 from \cite{Evans:2013}. Explicitly, in our calculations, the amplitude reflectivity $r_{\mathrm{fc}}(\Omega)$ of a two mirror filter cavity, for
an {\it audio sideband frequency} $\Omega$ is given by
\begin{equation}
    r_{\mathrm{fc}}(\Omega) = r_{\mathrm{in}} - \frac{t_{\mathrm{in}}^2}{r_{\mathrm{in}}} \frac{r_{\mathrm{rt}} e^{i\phi(\Omega)}}{1 - r_{\mathrm{rt}} e^{i\phi(\Omega)}},
    \label{eq:OFCrefl}
\end{equation}

where the symbols $r_{\rm rt}$, $r_{\rm in}$ and $\phi(\Omega)$ are defined in Eqs.\,A7--A8 of \cite{Evans:2013}.

Using the relation between quadrature fields $a_{1,2}(\Omega)$ and sideband fields $a(\omega_0\pm\Omega)$, Eq.~\eqref{eq:OFCrefl} allows us to calculate the phase rotation a quadrature field experiences upon reflection from the OFC, as

\begin{equation}
b_{\rm out} (\omega_0+\Omega)= r_{\rm fc}(\Omega)  b_{\rm in} (\omega_0+\Omega).
\end{equation}

\section{Effect of amplifier mirror mass on sensitivity}
\label{app:mirror_mass}

\begin{figure}
    \centering
    \includegraphics[width=\columnwidth]{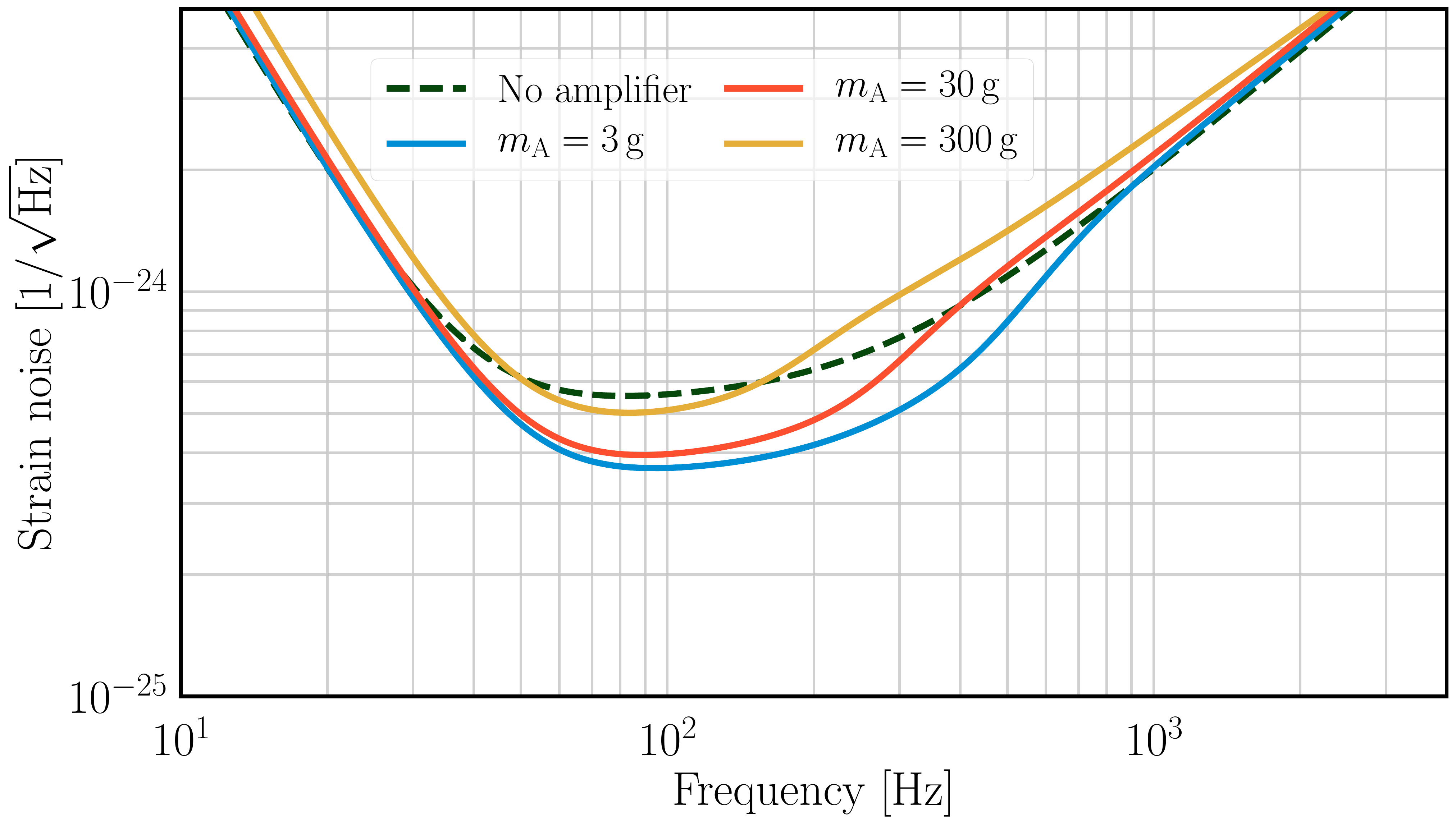}
    \caption{Voyager sensitivity improvement as a function of the mass of the amplifier mirrors. In all cases, 15\,dB of frequency-dependent squeezed vacuum is assumed to be injected into the anti-symmetric port of main interferometer.}
    \label{fig:massScaling}
\end{figure}

An important design consideration is the choice of mass of the amplifier mirrors. In~\Cref{fig:massScaling}, we show the sensitivity for three different mass choices (all for the case of 15\,dB frequency-dependent squeezed vacuum injection). The amplifier and OFC parameters for the 30\,g case are those given in~\Cref{tab:params}. However, for 3\,g and 300\,g, the amplifier properties (except for optical loss and displacement noise) and OFC properties (except for optical loss and cavity length) were re-adjusted to achieve optimal sensitivity. Lighter mirrors offer more sensitivity improvement, mainly due to the higher amplifier optomechanical gain. We therefore propose to make the mirror as light as possible.

However, there are several practical difficulties involved in working with very light mirrors. Firstly, with extremely light mirrors, it is difficult to sustain large circulating power in the amplifier ring cavities. Another concern is related to suspension thermal noise, which scales as $\propto m_{\rm A}^{-1/2}$. The difficulty in realizing low $\phi(f)$ for suspension fibers with a large surface-area-to-volume ratio motivates the choice of $m_{\rm A} = 30\,{\rm g}$. This allows for a wide margin of safety since the suspension noise is $\simeq 1/10$ of the limiting noises in our noise budget.

\section{Prospects for 20 dB squeeze injection}
\label{app:20dB}

\begin{figure}
    \centering
    \begin{subfigure}[t]{1\columnwidth}
        \includegraphics[width=\columnwidth]{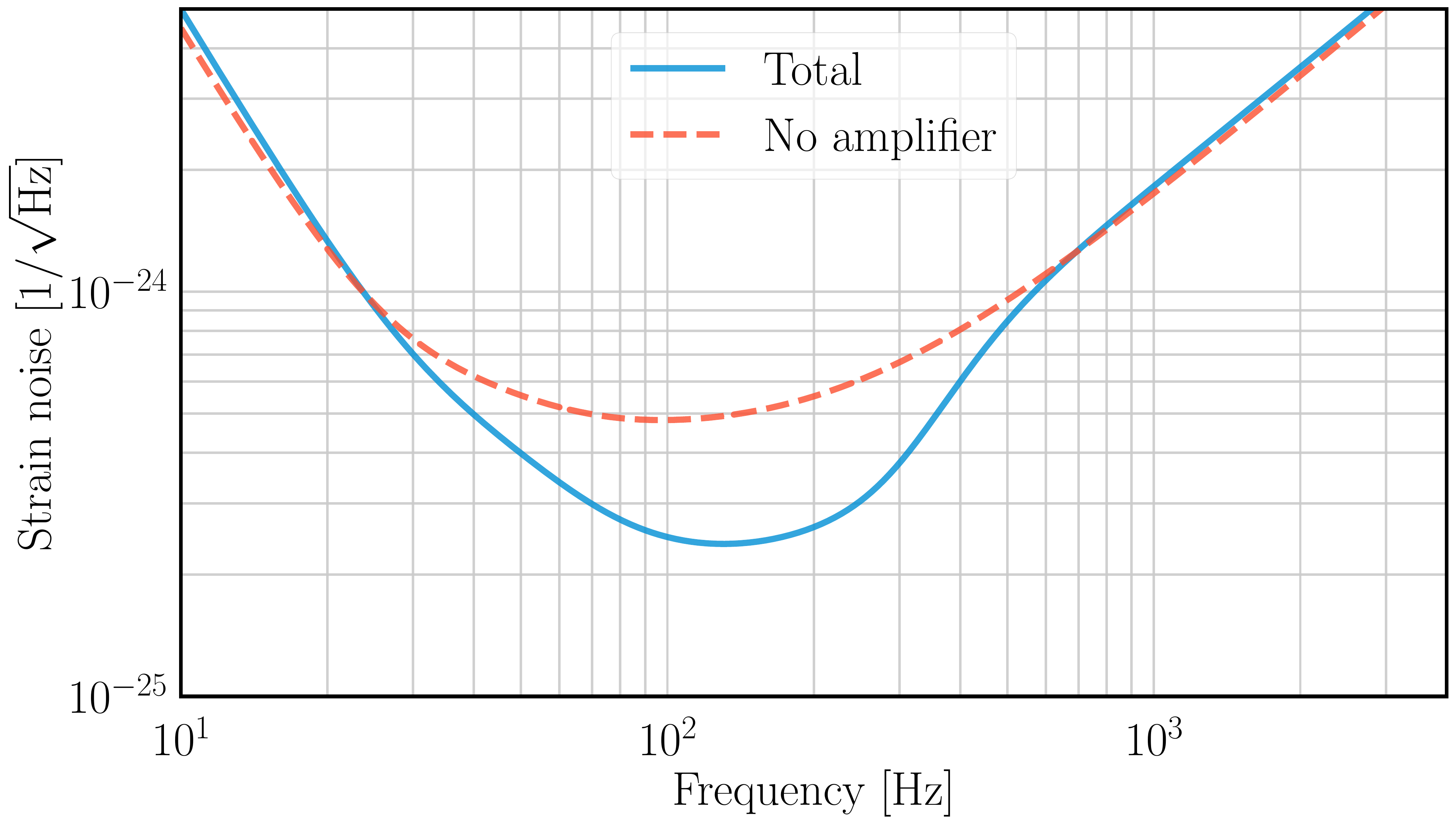}
    \end{subfigure}
    \begin{subfigure}[t]{1\columnwidth}
        \includegraphics[width=\columnwidth]{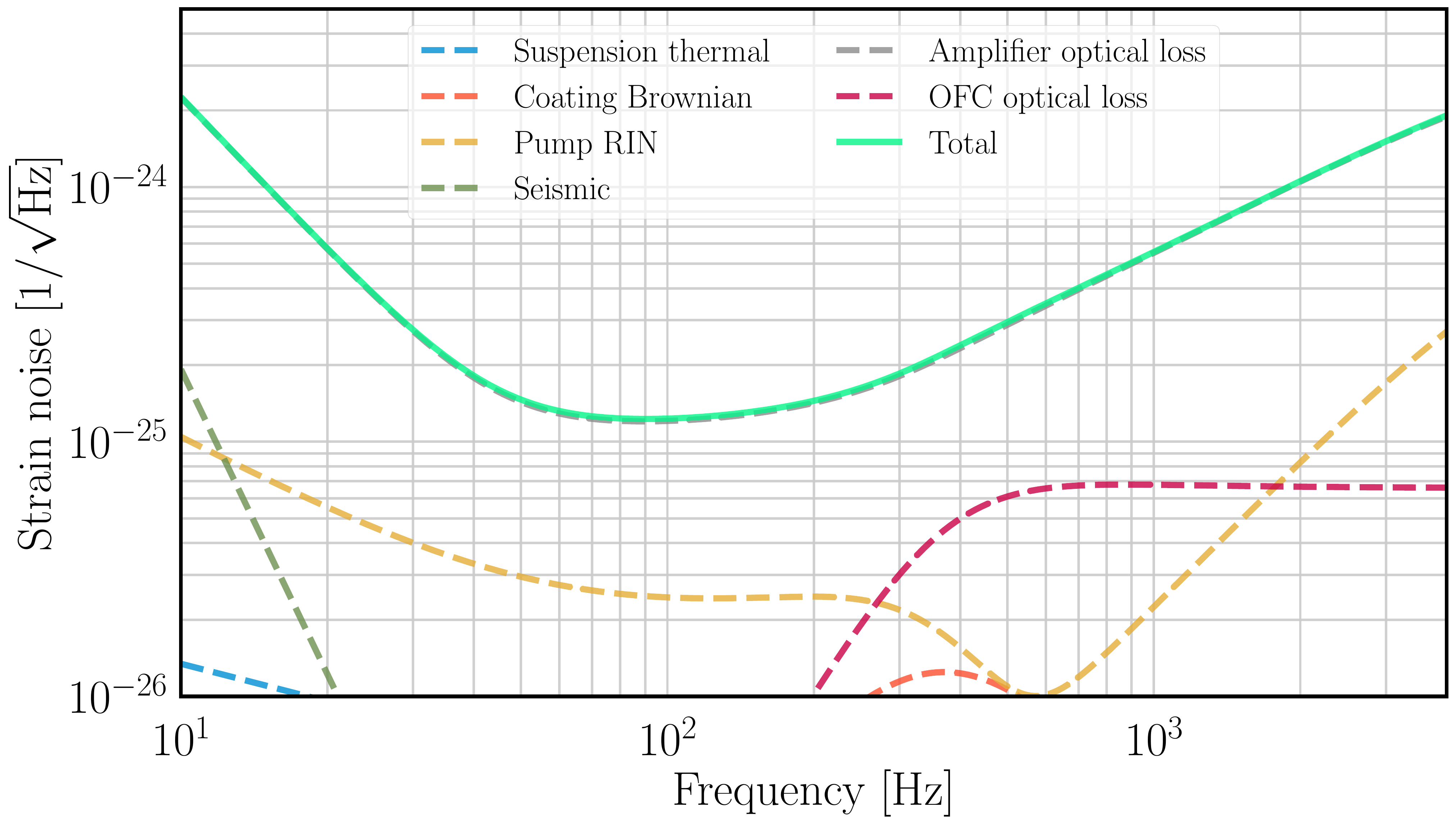}
    \end{subfigure}
    \begin{subfigure}[t]{1\columnwidth}
        \includegraphics[width=\columnwidth]{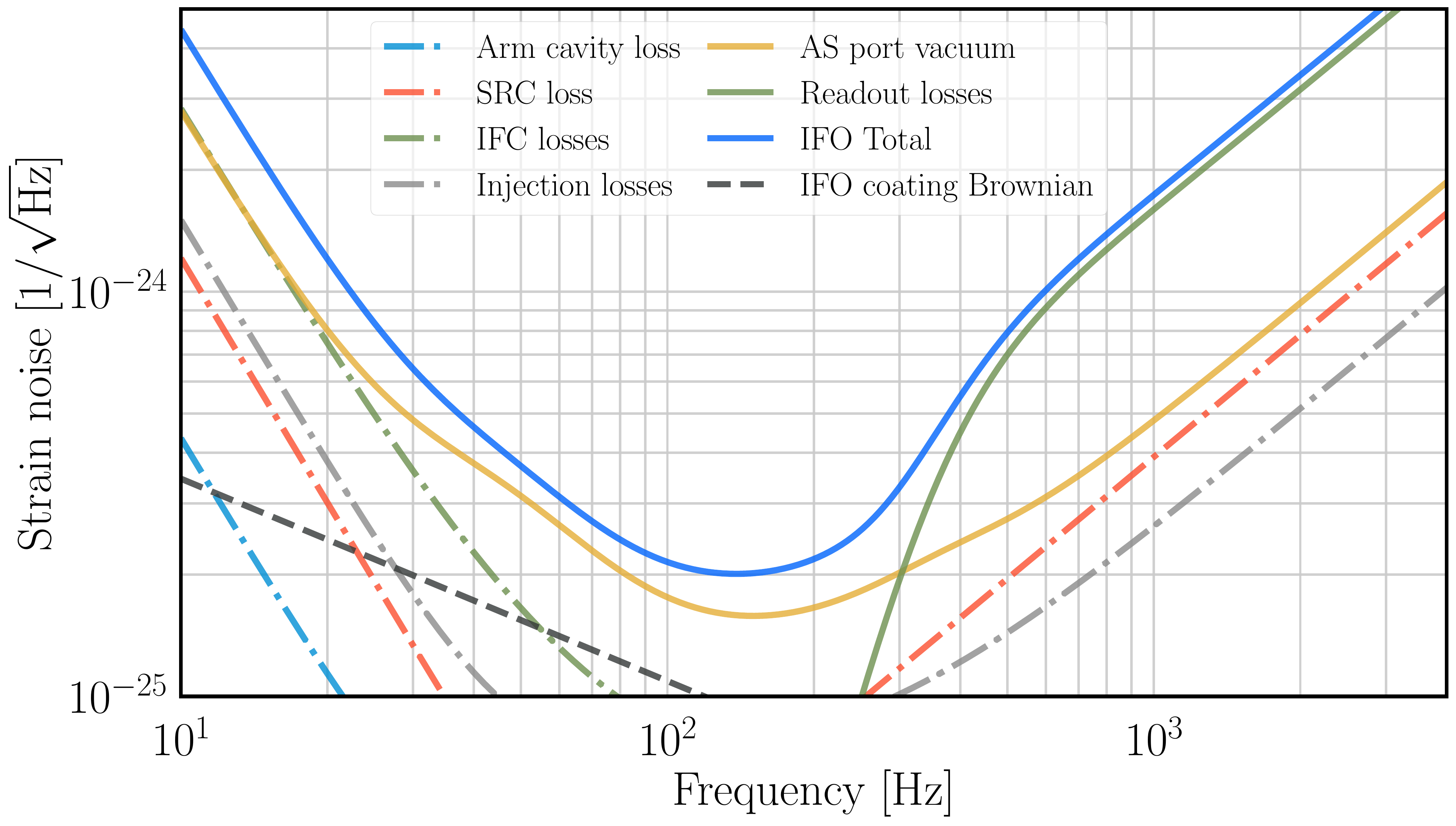}
    \end{subfigure}
    \caption{Sensitivity of LIGO Voyager for a more ambitious design that incorporates
      20\,dB frequency-dependent squeezed vacuum injection,
      and an amplifier with 10\,g mirrors.
      The detailed parameter assumptions are given in~\Cref{tab:params}. The interpretation of the top, middle, and bottom subplots are the same as the 15\,dB plots~\Cref{fig:total_budget},~\Cref{fig:amp_budget}, and~\Cref{fig:ifoNoises}, respectively. The coating Brownian noise shown in the lower plot is a factor of 5 lower than the nominal Voyager design.}
    \label{fig:speculative}
\end{figure}

We consider prospects for 20\,dB of frequency dependent squeezed vacuum injection, for which the amplifier makes a much bigger impact on the detector sensitivity. To achieve the correct rotation of the vacuum noise ellipse as a function of frequency, we find that two input filter cavities in series are needed, unlike the 15\,dB case, where only one is needed. Furthermore,
all of the noise sources within the main interferometer
must be maintained well below the squeezed vacuum noise, which would be extremely challenging at this level.
Finally, to fully take advantage of the sensitivity improvement, we propose using even lighter mirrors for the amplifier ($m_{\rm A}=10\, \rm g$), which results in higher optomechanical gain.
The detailed design parameters for all the optics and associated losses are given in~\Cref{tab:params}, and the corresponding noise curves are shown in~\Cref{fig:speculative}.

\clearpage

\bibliographystyle{unsrt}
\bibliography{gw_references}

\end{document}